\documentclass[sigconf]{acmart}

\AtBeginDocument{%
  }

\setcopyright{acmcopyright}
\copyrightyear{2023}
\acmYear{2023}
\acmDOI{XXXXXXX.XXXXXXX}

\acmBooktitle{CHI 2023 Workshop on HCI for Climate Change Imagining Sustainable Futures, ACM CHI Conference on Human Factors in Computing Systems, April 2023, Hamburg, Germany}
\usepackage{multirow}
\usepackage{comment}
\usepackage{natbib}
\usepackage{tabularx}
\usepackage{multicol, blindtext}

\makeatletter

\begin{document}

\title[Reflections on takeaway messages of climate change data visualizations]{``The main message is that sustainability would help'' -- Reflections on takeaway messages of climate change data visualizations}


    \author{Regina Schuster}
\email{regina.maria.veronika.schuster@univie.ac.at}
\affiliation{%
  \institution{University of Vienna}
  \country{Austria}
}

\author{Laura Koesten}
\email{laura.koesten@univie.ac.at}
\affiliation{%
  \institution{University of Vienna}
  \country{Austria}}

\author{Kathleen Gregory}
\email{kathleen.gregory@univie.ac.at}
\affiliation{%
  \institution{University of Vienna}
  \country{Austria}}

\author{Torsten Möller}
\email{torsten.moeller@univie.ac.at}
\affiliation{%
  \institution{University of Vienna}
  \country{Austria}}

\renewcommand{\shortauthors}{Schuster et al.}

\begin{abstract}
   How do different audiences make sense of climate change data visualizations and what do they take away as a main message? To investigate this question, we are building on the results of a previous study, focusing on expert opinions regarding public climate change communication and the role of data visualizations. Hereby, we conducted semi-structured interviews with $17$ experts in the fields of climate change, science communication, or data visualization. We also interviewed six lay persons with no professional background in either of these areas. With this analysis, we aim to shed light on how lay audiences arrive at an understanding of climate change data visualizations and what they take away as a main message. For two exemplary data visualizations, we compare their takeaway messages with messages formulated by experts. Through a thematic analysis, we observe differences regarding the included contents, the length and abstraction of messages, and the sensemaking process between and among the participant groups.
\end{abstract}




\maketitle

\section{Introduction} 
Although climate change has grown beyond a science and policy issue and has long entered the public discourse~\cite{moser_communicating_2010}, public attitudes towards climate change seem to ``have lagged behind the scientific evidence''~\cite{joslyn_explaining_2021}. Linking existing scientific knowledge to concrete action still appears to be a problem~\cite{naustdalslid_climate_2011, boehm_state_2021}, not least for lay people. With visual representations of data being crucial in communicating the science behind climate change~\cite{schneider_climate_2012, ipcc_new_2009}, communication practices are witnessing a trend toward visual formats~\cite{oneill_climate_2014, hansen_visually_2008, sheppard_can_2008} resulting in a growing interest by the HCI community~\cite{ferreira_climate_2021,ferreira_interacting_2021}. Visual data communication influences how people make decisions and can be used to shape opinions and attitudes~\cite{correll_ethical_2019} -- a crucial aspect for climate change communication. Yet, little is known about how different audiences make sense of climate change data visualizations~\cite{denniss_self_2015} and what they take away as a main message. With this analysis, we aim to shed light on how lay people arrive at an understanding of climate change data visualizations and how their takeaway messages might differ from messages formulated by experts. Hereby, we are building on the results of a previous study~\cite{Schuster2022BeingSO} for which we conducted semi-structured interviews with $17$ experts in the fields of climate change, science communication, or data visualization, as well as with six lay persons with no professional background in either of these areas. We used two exemplary climate change data visualizations as a discussion basis and asked participants about their personal main takeaway message. Through a thematic analysis, we observe differences regarding the included contents, the length and abstraction of messages, and the sensemaking process between and among the participant groups. 

\section{Research background} 
Human perception of data visualizations is guided by a variety of factors, including colors and patterns, which means that even design choices ``have the power to embolden a message or to mute it''~\cite{landers_storytelling_2019}. Such design choices and their influence on how people make sense of data visualizations have been studied from various angles reaching from the usage of pictographs~\cite{burns_designing_2022} to the visual arrangement of bar charts~\cite{xiong_visual_2021}. Besides the different foci, various approaches exist to investigate sensemaking processes (e.g. eye-tracking or mouse clicks~\cite{kim_bubbleview_2017}) as well as to test comprehension and understanding of specific data visualization types (e.g. posing questions to determine how well readers understand a visualization~\cite{canham_effects_2010, fischer_how_2018}). As an integral part of the sensemaking process, the formulation of takeaway messages can be used as another method for approximating readers' understanding of a data visualization~\cite{pinker_theory_1990, karer_conceptgraph_2020, burns2020how}. For example, \citet{bateman_useful_2010} tested the effect of `chart-junk' on the reader's comprehension and memorability and asked their study participants: `Is the author trying to communicate some message through the chart?'. They found that participants saw such messages significantly more often in charts with some sort of visual embellishments than in the plain versions. Another example of using takeaway messages in visual data communication in a climate change context is \citet{gammelgaard_ballantyne_images_2016}, who studied how students make sense of the messages conveyed through a dome theater movie about climate change. They found that the audience’s preconceptions of climate change influenced their interpretations of messages, potentially influencing their feeling of individual responsibility. In general, reader's takeaway messages can be seen as a personal summary of the key contents of a visualization. Naturally, the formulation of such messages is influenced by various factors including design choices and intended messages on the visualization producer's side, but also knowledge, experience and opinions on the reader's side. However, we know little about how different audiences arrive at an understanding of a data visualization and what contents they include in their personal takeaway message.

\section{Methodology} 

\begin{figure}

\begin{minipage}{.47\textwidth}
\centering
\resizebox{\textwidth}{!}{
    \begin{tabular}{l l l l}
        \toprule
        \textbf{ID} & \textbf{Age} & \textbf{Job role} & \textbf{Highest level of education} \\
        \midrule   
        L-1 & 18-24 & Care-giver apprentice & Intermediate school \\
        L-2 & 25-34 & Master student & University (Bachelor degree) \\
        L-3 & 35-44 & Clerk in accounting & Grammar school \\
        L-4 & 35-44 & Dentist & University (Doctoral degree) \\
        L-5 & 55-64 & Retired machine fitter & Secondary general school \\
        L-6 & 65-74 & House-wife & Secondary general school \\
        \bottomrule
    \end{tabular}}
    \captionof{table}{Participants' self-reported age group, job role, and highest level of education}\label{tab:participants}
\resizebox{\textwidth}{!}{
    \begin{tabular}{l | c c c c c c | c}
        \toprule
        \textbf{News media} & \textbf{L-1} & \textbf{L-2} & \textbf{L-3} & \textbf{L-4} & \textbf{L-5} & \textbf{L-6} & \textbf{SUM}\\
        \midrule   
           TV &                 &     x &     x &     x &     x &     x &      5 \\
           Radio &            x &     x &       &       &     x &     x &      4 \\
           Online &           x &     x &     x &     x &       &       &      4 \\
           Social media &     x &     x &     x &     x &       &       &      4 \\
           Print &              &       &       &     x &       &     x &      2 \\
        \bottomrule
    \end{tabular}}
    \captionof{table}{Participants' self-reported used news media sources (x indicates that a source is used by a participant)}\label{tab:newsmedia}
\end{minipage}
\hspace{1cm}
\begin{minipage}{.47\textwidth}
    \centering
    \includegraphics[width=1\linewidth]{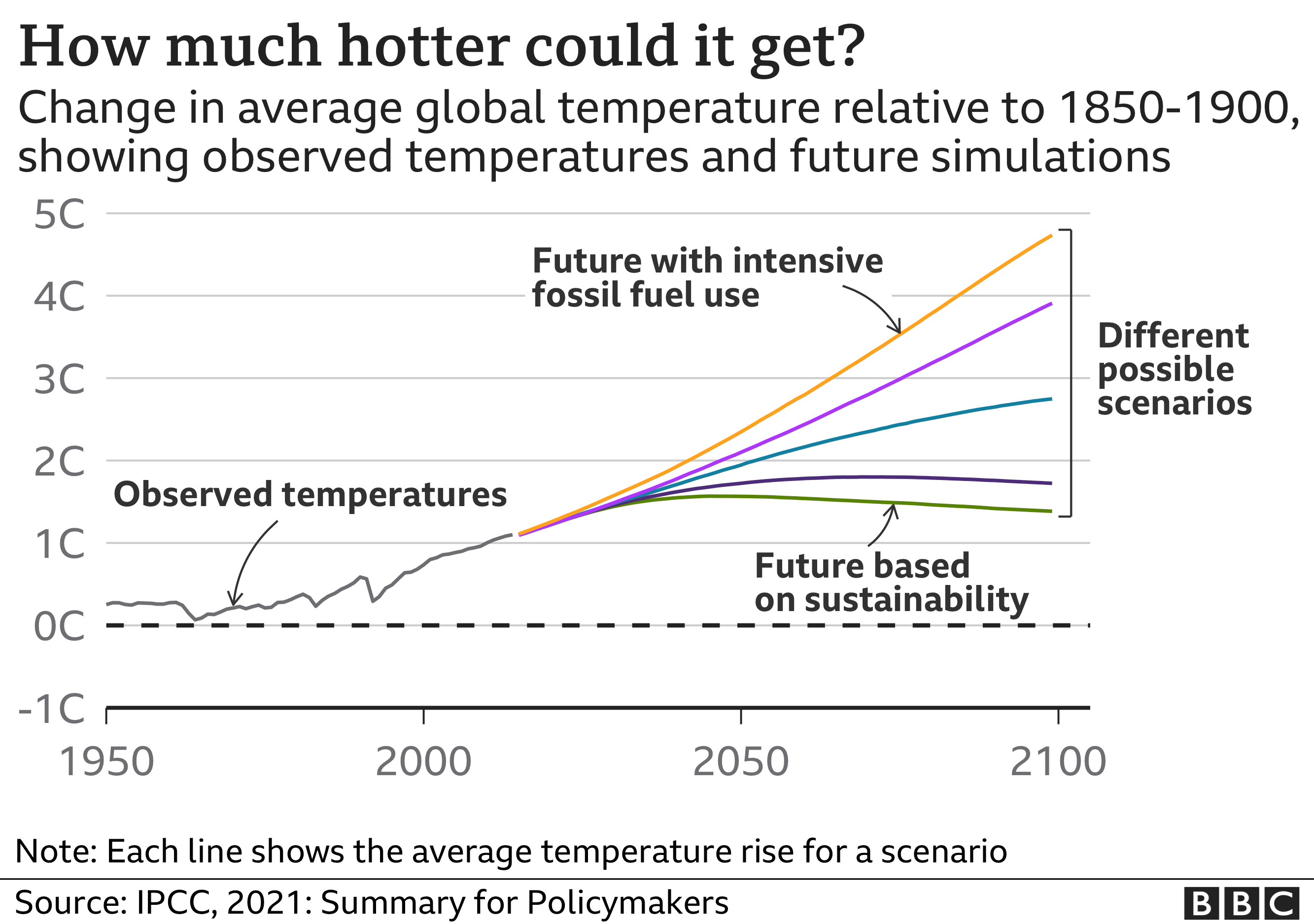}
    \captionof{figure}{Example visualization 1 created by and reproduced with permission of BBC News at bbc.co.uk/news \cite{BBC}}
    \label{fig:bbc}
\end{minipage}%
\end{figure}

As part of a larger project investigating expert opinions on public climate change communication and the role of data visualizations~\cite{Schuster2022BeingSO}, we conducted semi-structured interviews with $17$ experts in the fields of climate change, science communication, or data visualization. We also interviewed six members of the general public with no professional background in either of these areas, who we refer to as lay participants. In the following, lay participants are identified as L-1 to L-6 and expert participants as E-1 to E-17. The focus of this analysis lies on the lay interview participants, who were selected on the basis of diversity factors regarding age, gender, education, and occupation. Table~\ref{tab:participants} shows their self-reported age group, job role, and education; Table~\ref{tab:newsmedia} shows the sources participants use for news consumption. All lay participants were German or Austrian residents. The interviews were held in-person or via Zoom according to the participants' preferences and included questions related to participants' opinions and experiences with (visual) climate change communication. We used two example visualizations from online news sources (\autoref{fig:bbc} and a figure published by The Guardian~\cite{the_guardian_as_2021}; if preferred, translated to German) as a discussion basis and asked participants about the main takeaway message the visualization author wanted to convey. To be able to compare results, we posed the same question to up to $17$ expert participants depending on time during the interview. After transcription, we conducted a thematic analysis using the qualitative data analysis software Atlas.ti. We created a codebook by systematically going through the transcriptions employing a combination of deductive and inductive analysis~\cite{robson2017real}. Broader themes that make up the results and supporting categories of codes were developed according to \citet{vaismoradi_theme_2016}. More details about the methodology, the choice of example visualizations, the interview procedure, and the expert participants are detailed in our previous work~\cite{Schuster2022BeingSO}.

\section{Results} 
To assess the lay participants' attitudes towards the topic, we asked them if climate change is something they think about often, which all six participants confirmed. In explanation, they named the feeling that they are or will be affected by climate change and that we urgently need to act as a society and as individuals. Participants also expressed anger towards politicians, restricting regulations, and the perceived lack of action. Some participants argued that climate change is important to them because of the (extensive) reporting of climate issues in the news media. Others mentioned that climate change is not covered enough in classic and non-classic news sources (e.g., leisure journals and social media) and that climate change phenomena are usually not explained well enough.

To learn more about the lay participants' opinions regarding climate change data visualizations, the majority of the discussion revolved around the two example figures, in particular their attractiveness, understandability, and perceived takeaway messages. \autoref{fig:bbc} was also shown to $17$ experts, resulting in a total of $23$ formulated takeaway messages. Example visualization 2~\cite{the_guardian_as_2021} was shown to all six lay participants and to seven experts, resulting in a total of $13$ messages. For both example visualizations and an overall count of $36$ formulated takeaway messages for thematic analysis, we have observed variations in the included contents and the length/abstraction of messages, as well as in the sensemaking process itself among the two participant groups, but also between lay and expert participants.

\subsection{Contents of the messages} 
As a first step in the message analysis, we looked at the message contents, i.e. the subjects or topics that lay and expert participants have included in their messages. Codes for the contents were created in an iterative process according to their usage in messages. For example, the codes for the subjects `temperature' and `increase' were separated, as not all messages included both of these thematic concepts. Looking at the mentioned contents in the messages, we can see a high variation in what participants regard as the main takeaway of a visualization. Both lay and expert participants formulated diverse takeaway messages, with some not even mentioning temperature. Due to the similar nature of the visualizations, the message contents of both visualizations can be organized into three broader categories: content regarding temperature, effects and influences, and time.

\textbf{Example visualization 1.} For \autoref{fig:bbc}, concrete subjects that are apparent in more than half of the messages by lay and expert participants include future, positive/negative influences or effects, global earth average, temperature, increase, and scenarios. Some aspects, like the possibility of a temperature decrease in a very optimistic scenario, fossil fuel use, and sustainability, were mentioned more often by lay participants than experts. The positive/negative influence or effect of certain policies was more frequently included in the experts' messages. \autoref{tab:contents} shows example messages formulated by lay and expert participants and the respective coded contents.

\textbf{Example visualization 2.} For example visualization 2~\cite{the_guardian_as_2021}, concrete subjects apparent in more than half of the messages included the reference to temperature and statements about an increase. Lay participants mentioned the color red and specifics about the line in the chart as part of their messages more often than experts. \autoref{tab:contents} shows example messages emphasizing the difference in mentioned contents between lay and expert participants.

\begin{table*}[ht]
\small
\centering
\begin{tabularx}{\textwidth}{p{0.65\textwidth}p{0.3\textwidth}}

    \hline
    \textbf{Example messages} & \textbf{Coded contents} \\
    \midrule 
    
    ``The effect that sustainability could have on the temperature or global temperature on average. And there are like 1, 2, 3, 4, 5 scenarios. So I think the main message is that sustainability would help. And that if you continue to live like we do it, we would be at plus five, I guess degree Celsius.'' (L-2 about example visualization 1) 
    & Sustainability, positive/negative influence or effect, global earth average, temperature, scenarios, 5-degrees \\
    \multicolumn{1}{c}{ }&\multicolumn{1}{c}{ }\\[-0.2cm]
    
    ``They show how different the future can be, depending on humans, how humans behave.'' (E-14 about example visualization 1) 
    & Positive/negative influence or effect, future, human influence or fault \\
    \multicolumn{1}{c}{ }&\multicolumn{1}{c}{ }\\[-0.2cm]

    ``Things are going extremely up now. Like the red line, from 1850 it was still down, and then in 2020 we are already a long way up.'' (L-6 about example visualization 2)  
    & Increase, severity, red, line, year-1850, year-2020, difference \\
    \multicolumn{1}{c}{ }&\multicolumn{1}{c}{ }\\[-0.2cm]

    ``We see that in a very short time that the Earth got hotter. So there's no way this is a normal cycle if you compare it to the historic data. So it's us basically, is what the message is.'' (E-13 about example visualization 2)
    & Short time frame, global earth average, temperature, increase, anomaly, past, human influence or fault \\
    
    \hline
\end{tabularx}
\captionof{table}{Example messages and coded contents}\label{tab:contents}
\vspace{-7mm}
\end{table*}

\subsection{Length, detail, and abstraction}  
For both example visualizations, the median length of formulated takeaway messages is more than double for the lay participants compared to the experts' messages. A reason for that might lie in the different levels of abstraction that are common among these two groups. While experts tended to formulate shorter, more abstract messages, lay participants focused more on concrete facts and very specific details.

\textbf{Example visualization 1.} With a median length of $73$ words per message, messages formulated by lay participants are considerably longer than the experts' messages (median of $29$ words per message). Concrete temperature specifications or references to specific years were commonly included in lay participants' messages, but not mentioned in any of the experts' messages. Overall, interview partners from the lay group focused more on stating facts or details displayed in the visualization, while experts formulated more abstract messages, often calling on our ability to shape the future, as can be seen in the following example messages: \textit{``If we focus on sustainability in the future, the global average temperature can be kept below the two-degree target that is being striven for, or even below 1.5 degrees and then even drop slightly by 2100. And in contrast, if we keep using everything fossil the way we have been using it, by 2100 we will be close to five degrees of global warming.''} (L-4) vs. \textit{``How much possibility mankind has to influence how the average global temperature will be in the next 50 years.''} (E-8).

\textbf{Example visualization 2.} We again see a longer median length in messages by lay participants than by experts (median of $60$ vs. $28$ words per message). While lay participants more frequently referred to specific years, experts used more general time frames to address events in the past or the present. The difference in the level of abstraction between lay and expert messages can be seen in the following example: \textit{``It's clearly about the global surface temperature, again viewed over different years. And the main message is now, if I look at 2020 for example, I see that the red line shoots up here, and steeply at that. [...] A retrospect, where it was actually, I would say, relatively the same and then a little lower and then it got higher again and now it is shooting up.''} (L-3) vs. \textit{``Climate change is human-driven.''} (E-11).

\subsection{Understandability and sensemaking}  
Some aspects of the message formulation and the subsequent discussion of the visualizations gave hints about whether or not the participants interpreted the graphs in a way that supports the shown data. As an optional task, lay participants were prompted to share their thinking process out loud while looking at the visualizations for the first time. Hence in some cases, participants named the chart parts they were looking at one by one. Hereby, we saw that sensemaking processes also seem to be highly individual: while some read the textual parts, like the title/subtitle, before looking at the graph, others first explored the different parts of the graph itself and only afterward the text. For both visualizations, some participants mentioned that their attention was guided by the color of the objects. It stands out that contrary to experts, lay participants used formulations close or identical to the text on the visualizations. Experts not only tended to abstract the takeaway messages but also arrived at their conclusions faster, presumably because most of them appeared to have seen versions of the graphs before. For the lay participants, attitudes toward their ability to read data visualizations and their general interest in the topic seemed to influence their efforts during the message formulation.  

\textbf{Example visualization 1.} Four lay participants appeared to interpret the visualization in a way that matched our interpretation as researchers; two actively expressed that they were having difficulties. One of them (L-1) voiced statements that make us doubt a correct interpretation: \textit{``This fuel and sustainability actually have nothing to do with the rise in temperature and anything like that. [...] I think young and old know that if there's a minus in front, it's lousy.''} (L-1). Four participants assessed the interpretation of \autoref{fig:bbc} as being too difficult to understand for lay people (in general, not necessarily themselves). Another four lay participants stated that \autoref{fig:bbc} would not catch their attention in their day-to-day news consumption. While half of the participants thought that the creator of \autoref{fig:bbc} overall did a good job of conveying the main message, two participants felt that the creator did not adequately convey the message.

\textbf{Example visualization 2.} The same two participants, who expressed difficulties with interpreting the first visualization, did also do so for the second one. Three lay participants appeared to interpret the graph in a way that is supported by the shown data, while the interpretation of the other three was unclear. Two participants explicitly cited the title of the visualization in their takeaway message. However, not only members of the lay group had difficulties with the chart, but also two experts stated that at first sight, they would struggle to derive a correct interpretation: \textit{``It isn't very clear. Something happened past 1850 and then in the beginning of the 19th century.''} (E-8). Four lay participants assessed the visualization as being easy to interpret for lay people, while two thought it is too difficult. Another four participants stated that the figure would catch their attention in their daily news consumption. One person thought the creator of this visualization did a bad job of conveying the main message, mainly because they perceived the visualization as not trustworthy; however, the majority of participants (four) felt that the creator did convey the message in a good way.

\section{Discussion} 
We investigated $36$ takeaway messages formulated by lay and expert interview participants about two climate change data visualizations. Similar findings were apparent for both visualizations, such as the individuality of the sensemaking process and the included contents, as well as the difference in length and abstraction for lay vs. expert messages. 

\textbf{How did participants make sense of climate change data visualizations?} 
In accordance with previous literature (e.g.~\cite{Deng2016, denniss_self_2015, rothermich_influence_2021, nadeem_gen_2021}), our results suggest that the process of making sense of data visualizations is highly individual and influenced by a person's previous knowledge, attitudes and opinions. The steps taken to make sense of the visualizations varied between lay persons with different features of the graph attracting the attention of different persons. Especially for the lay participants, the color of objects and textual parts of the figures seem to have influenced their attention and subsequently the formulation of their takeaway message. Participants who expressed that they did not feel equipped to interpret data visualizations and lack knowledge about climate change, have shown less effort to engage with the figures. Contrarily, participants with a high interest in the topic took more time and effort to interpret the data visualizations. We also noticed that experts seemed to arrive at a main message faster than the lay group. This might be due to familiarity with the topic in general as well as with the particular contents of the visualizations. Lay people also differed in their assessment of the understandability of the example visualizations with participants from various backgrounds finding the example visualizations too difficult to interpret for lay audiences.

\textbf{What did participants take away as a main message from a visualization?} 
Besides the sensemaking process, also contents, length, and abstraction of the takeaway messages highly varied amongst participants. Aspects that are essential to one person, get ignored by another, which might eventually shape what a person takes away and remembers from a data visualization~\cite{bateman_useful_2010}. We also found a difference between lay and expert participants for the derived takeaway messages: experts' messages were shorter and more abstract. Despite the different application in natural language descriptions of visualizations, our findings regarding the content of takeaway messages can be compared to the four levels of semantic content by~\citet{lundgard_accessible_2021}, namely 1) visualization construction properties, 2) statistical concepts and relations, 3) perceptual and cognitive phenomena, and 4) domain-specific insights. While lay participants tended to include lower level contents in their messages referring to the elements of the visualization or statistical details (level 1 and 2 according to~\cite{lundgard_accessible_2021}), expert participants more frequently used higher level contents like trends, explanations or social/political context (level 3 and 4 according to~\cite{lundgard_accessible_2021}). However, not only the participants' background but also the design of the visualization itself seems to have influenced the semantic content of the takeaway messages. A noticeable difference between the two example visualizations lies in the references of visualization properties (level 1 according to~\cite{lundgard_accessible_2021}): neither lay nor expert participants have mentioned concrete visualization parts, like colors or lines, in their messages for the first example visualization but commonly did so for the second one. This might be due to the bolder design choices apparent in the visualization, which even influenced the perceived trustworthiness.

\textbf{Limitations and future work.} 
As with any kind of research, we have to understand our findings within the limitations of this study setup. With this qualitative approach and a sample of six interview participants, we cannot claim generalizability for our results. Rather, they can be seen as an impulse for future work directions. Influencing factors on how persons formulate takeaway messages for climate change data visualizations could be explored further with a larger and a more diverse (culturally and geographically) sample. We acknowledge that the chosen example visualizations have shaped our results, which could be accounted for with different chart types in follow-up studies. With future projects testing aspects of the understandability of data visualizations, we hope to contribute further to existing knowledge in the field of visual climate change communication.
\begin{acks}
  This project has been funded by the Vienna Science and Technology Fund (WWTF) [10.47379/ICT20065].
\end{acks}
\bibliographystyle{ACM-Reference-Format}
\bibliography{bibliography}


\begin{thebibliography}{33}


\ifx \showCODEN    \undefined \def \showCODEN     #1{\unskip}     \fi
\ifx \showDOI      \undefined \def \showDOI       #1{#1}\fi
\ifx \showISBNx    \undefined \def \showISBNx     #1{\unskip}     \fi
\ifx \showISBNxiii \undefined \def \showISBNxiii  #1{\unskip}     \fi
\ifx \showISSN     \undefined \def \showISSN      #1{\unskip}     \fi
\ifx \showLCCN     \undefined \def \showLCCN      #1{\unskip}     \fi
\ifx \shownote     \undefined \def \shownote      #1{#1}          \fi
\ifx \showarticletitle \undefined \def \showarticletitle #1{#1}   \fi
\ifx \showURL      \undefined \def \showURL       {\relax}        \fi
\providecommand\bibfield[2]{#2}
\providecommand\bibinfo[2]{#2}
\providecommand\natexlab[1]{#1}
\providecommand\showeprint[2][]{arXiv:#2}

\bibitem[Ballantyne et~al\mbox{.}(2016)]%
        {gammelgaard_ballantyne_images_2016}
\bibfield{author}{\bibinfo{person}{Anne~Gammelgaard Ballantyne},
  \bibinfo{person}{Victoria Wibeck}, {and} \bibinfo{person}{Tina-Simone
  Neset}.} \bibinfo{year}{2016}\natexlab{}.
\newblock \showarticletitle{Images of climate change – {A} pilot study of
  young people’s perceptions of {ICT}-based climate visualization}.
\newblock \bibinfo{journal}{\emph{Climatic Change}} \bibinfo{volume}{134},
  \bibinfo{number}{1} (\bibinfo{date}{Jan.} \bibinfo{year}{2016}),
  \bibinfo{pages}{73--85}.
\newblock
\showISSN{1573-1480}
\urldef\tempurl%
\url{https://doi.org/10.1007/s10584-015-1533-9}
\showDOI{\tempurl}


\bibitem[Bateman et~al\mbox{.}(2010)]%
        {bateman_useful_2010}
\bibfield{author}{\bibinfo{person}{Scott Bateman}, \bibinfo{person}{Regan~L.
  Mandryk}, \bibinfo{person}{Carl Gutwin}, \bibinfo{person}{Aaron Genest},
  \bibinfo{person}{David McDine}, {and} \bibinfo{person}{Christopher Brooks}.}
  \bibinfo{year}{2010}\natexlab{}.
\newblock \showarticletitle{Useful Junk? The Effects of Visual Embellishment on
  Comprehension and Memorability of Charts}. In
  \bibinfo{booktitle}{\emph{Proceedings of the SIGCHI Conference on Human
  Factors in Computing Systems}} (Atlanta, Georgia, USA)
  \emph{(\bibinfo{series}{CHI '10})}. \bibinfo{publisher}{Association for
  Computing Machinery}, \bibinfo{address}{New York, NY, USA},
  \bibinfo{pages}{2573–2582}.
\newblock
\showISBNx{9781605589299}
\urldef\tempurl%
\url{https://doi.org/10.1145/1753326.1753716}
\showDOI{\tempurl}


\bibitem[Boehm et~al\mbox{.}(2021)]%
        {boehm_state_2021}
\bibfield{author}{\bibinfo{person}{Sophie Boehm}, \bibinfo{person}{Katie
  Lebling}, \bibinfo{person}{Kelly Levin}, \bibinfo{person}{Hanna Fekete},
  \bibinfo{person}{Joel Jaeger}, \bibinfo{person}{Richard Waite},
  \bibinfo{person}{Anna Nilsson}, \bibinfo{person}{Joe Thwaites},
  \bibinfo{person}{Ryan Wilson}, \bibinfo{person}{Andreas Geiges},
  \bibinfo{person}{Clea Schumer}, \bibinfo{person}{Maggie Dennis},
  \bibinfo{person}{Katie Ross}, \bibinfo{person}{Sebastian Castellanos},
  \bibinfo{person}{Rajat Shrestha}, \bibinfo{person}{Neelam Singh},
  \bibinfo{person}{Mikaela Weisse}, \bibinfo{person}{Leah Lazer},
  \bibinfo{person}{Louise Jeffery}, \bibinfo{person}{Lydia Freehafer},
  \bibinfo{person}{Erin Gray}, \bibinfo{person}{Lihuan Zhou},
  \bibinfo{person}{Matthew Gidden}, {and} \bibinfo{person}{Madeleine Galvin}.}
  \bibinfo{year}{2021}\natexlab{}.
\newblock \bibinfo{title}{State of climate action 2021: {S}ystems
  transformations required to limit global warming to 1.5°{C}}.
\newblock
\newblock
\urldef\tempurl%
\url{https://doi.org/10.46830/wrirpt.21.00048}
\showDOI{\tempurl}


\bibitem[Burns et~al\mbox{.}(2020)]%
        {burns2020how}
\bibfield{author}{\bibinfo{person}{Alyxander Burns}, \bibinfo{person}{Cindy
  Xiong}, \bibinfo{person}{Steven Franconeri}, \bibinfo{person}{Alberto Cairo},
  {and} \bibinfo{person}{Narges Mahyar}.} \bibinfo{year}{2020}\natexlab{}.
\newblock \showarticletitle{How to evaluate data visualizations across
  different levels of understanding}. In \bibinfo{booktitle}{\emph{2020 IEEE
  Workshop on Evaluation and Beyond - Methodological Approaches to
  Visualization (BELIV)}}. \bibinfo{pages}{19--28}.
\newblock
\urldef\tempurl%
\url{https://doi.org/10.1109/BELIV51497.2020.00010}
\showDOI{\tempurl}


\bibitem[Burns et~al\mbox{.}(2022)]%
        {burns_designing_2022}
\bibfield{author}{\bibinfo{person}{Alyxander Burns}, \bibinfo{person}{Cindy
  Xiong}, \bibinfo{person}{Steven Franconeri}, \bibinfo{person}{Alberto Cairo},
  {and} \bibinfo{person}{Narges Mahyar}.} \bibinfo{year}{2022}\natexlab{}.
\newblock \showarticletitle{Designing with pictographs: {E}nvision topics
  without sacrificing understanding}.
\newblock \bibinfo{journal}{\emph{IEEE Transactions on Visualization and
  Computer Graphics}} \bibinfo{volume}{28}, \bibinfo{number}{12}
  (\bibinfo{date}{Dec.} \bibinfo{year}{2022}), \bibinfo{pages}{4515--4530}.
\newblock
\showISSN{1941-0506}
\urldef\tempurl%
\url{https://doi.org/10.1109/TVCG.2021.3092680}
\showDOI{\tempurl}


\bibitem[Canham and Hegarty(2010)]%
        {canham_effects_2010}
\bibfield{author}{\bibinfo{person}{Matt Canham} {and} \bibinfo{person}{Mary
  Hegarty}.} \bibinfo{year}{2010}\natexlab{}.
\newblock \showarticletitle{Effects of knowledge and display design on
  comprehension of complex graphics}.
\newblock \bibinfo{journal}{\emph{Learning and Instruction}}
  \bibinfo{volume}{20}, \bibinfo{number}{2} (\bibinfo{date}{April}
  \bibinfo{year}{2010}), \bibinfo{pages}{155--166}.
\newblock
\showISSN{0959-4752}
\urldef\tempurl%
\url{https://doi.org/10.1016/j.learninstruc.2009.02.014}
\showDOI{\tempurl}


\bibitem[Correll(2019)]%
        {correll_ethical_2019}
\bibfield{author}{\bibinfo{person}{Michael Correll}.}
  \bibinfo{year}{2019}\natexlab{}.
\newblock \showarticletitle{Ethical dimensions of visualization research}. In
  \bibinfo{booktitle}{\emph{Proceedings of the 2019 {CHI} {Conference} on
  {Human} {Factors} in {Computing} {Systems}}}. \bibinfo{publisher}{ACM},
  \bibinfo{address}{Glasgow Scotland Uk}, \bibinfo{pages}{1--13}.
\newblock
\showISBNx{978-1-4503-5970-2}
\urldef\tempurl%
\url{https://doi.org/10.1145/3290605.3300418}
\showDOI{\tempurl}


\bibitem[Deng and Sloutsky(2016)]%
        {Deng2016}
\bibfield{author}{\bibinfo{person}{Sophia Deng} {and} \bibinfo{person}{Vladimir
  Sloutsky}.} \bibinfo{year}{2016}\natexlab{}.
\newblock \showarticletitle{Selective attention, diffused attention, and the
  development of categorization}.
\newblock \bibinfo{journal}{\emph{Cognitive Psychology}}  \bibinfo{volume}{91}
  (\bibinfo{date}{12} \bibinfo{year}{2016}).
\newblock
\urldef\tempurl%
\url{https://doi.org/10.1016/j.cogpsych.2016.09.002}
\showDOI{\tempurl}


\bibitem[Denniss and Davison(2015)]%
        {denniss_self_2015}
\bibfield{author}{\bibinfo{person}{Rebecca~Joy Denniss} {and}
  \bibinfo{person}{Aidan Davison}.} \bibinfo{year}{2015}\natexlab{}.
\newblock \showarticletitle{Self and world in lay interpretations of climate
  change}.
\newblock \bibinfo{journal}{\emph{International Journal of Climate Change
  Strategies and Management}} \bibinfo{volume}{7}, \bibinfo{number}{2}
  (\bibinfo{date}{Jan.} \bibinfo{year}{2015}), \bibinfo{pages}{140--153}.
\newblock
\showISSN{1756-8692}
\urldef\tempurl%
\url{https://doi.org/10.1108/IJCCSM-03-2014-0046}
\showDOI{\tempurl}


\bibitem[Ferreira et~al\mbox{.}(2021a)]%
        {ferreira_climate_2021}
\bibfield{author}{\bibinfo{person}{Marta Ferreira}, \bibinfo{person}{Miguel
  Coelho}, \bibinfo{person}{Valentina Nisi}, {and} \bibinfo{person}{Nuno
  Jardim~Nunes}.} \bibinfo{year}{2021}\natexlab{a}.
\newblock \showarticletitle{Climate change communication in {HCI}: {A} visual
  analysis of the past decade}. In \bibinfo{booktitle}{\emph{Creativity and
  {Cognition}}}. \bibinfo{publisher}{ACM}, \bibinfo{address}{Virtual Event
  Italy}, \bibinfo{pages}{1--16}.
\newblock
\showISBNx{978-1-4503-8376-9}
\urldef\tempurl%
\url{https://doi.org/10.1145/3450741.3466774}
\showDOI{\tempurl}


\bibitem[Ferreira et~al\mbox{.}(2021b)]%
        {ferreira_interacting_2021}
\bibfield{author}{\bibinfo{person}{Marta Ferreira}, \bibinfo{person}{Nuno
  Nunes}, {and} \bibinfo{person}{Valentina Nisi}.}
  \bibinfo{year}{2021}\natexlab{b}.
\newblock \showarticletitle{Interacting with climate change: {A} survey of
  {HCI} and design projects and their use of transmedia storytelling}. In
  \bibinfo{booktitle}{\emph{Interactive {Storytelling}}}
  \emph{(\bibinfo{series}{Lecture {Notes} in {Computer} {Science}})},
  \bibfield{editor}{\bibinfo{person}{Alex Mitchell} {and}
  \bibinfo{person}{Mirjam Vosmeer}} (Eds.). \bibinfo{publisher}{Springer
  International Publishing}, \bibinfo{address}{Cham},
  \bibinfo{pages}{338--348}.
\newblock
\showISBNx{978-3-030-92300-6}
\urldef\tempurl%
\url{https://doi.org/10.1007/978-3-030-92300-6_33}
\showDOI{\tempurl}


\bibitem[Fischer et~al\mbox{.}(2018)]%
        {fischer_how_2018}
\bibfield{author}{\bibinfo{person}{Helen Fischer}, \bibinfo{person}{Stefanie
  Schütte}, \bibinfo{person}{Anneliese Depoux}, \bibinfo{person}{Dorothee
  Amelung}, {and} \bibinfo{person}{Rainer Sauerborn}.}
  \bibinfo{year}{2018}\natexlab{}.
\newblock \showarticletitle{How well do {COP22} attendees understand graphs on
  climate change health impacts from the {Fifth} {IPCC} {Assessment} {Report}?}
\newblock \bibinfo{journal}{\emph{International Journal of Environmental
  Research and Public Health}} \bibinfo{volume}{15}, \bibinfo{number}{5}
  (\bibinfo{date}{May} \bibinfo{year}{2018}), \bibinfo{pages}{875}.
\newblock
\showISSN{1660-4601}
\urldef\tempurl%
\url{https://doi.org/10.3390/ijerph15050875}
\showDOI{\tempurl}


\bibitem[Giuseffi et~al\mbox{.}(2019)]%
        {landers_storytelling_2019}
\bibfield{author}{\bibinfo{person}{Karl Giuseffi}, \bibinfo{person}{Benjamin
  Sievert}, \bibinfo{person}{Brett~M. Wells}, {and} \bibinfo{person}{Fran
  Westfall}.} \bibinfo{year}{2019}\natexlab{}.
\newblock \showarticletitle{Storytelling and sensemaking through data
  visualization}.
\newblock In \bibinfo{booktitle}{\emph{The {Cambridge} {Handbook} of
  {Technology} and {Employee} {Behavior}}},
  \bibfield{editor}{\bibinfo{person}{Richard~N. Landers}} (Ed.).
  \bibinfo{publisher}{Cambridge University Press},
  \bibinfo{address}{Cambridge}, \bibinfo{pages}{836--846}.
\newblock
\showISBNx{978-1-108-47670-6}
\urldef\tempurl%
\url{https://doi.org/10.1017/9781108649636.031}
\showDOI{\tempurl}


\bibitem[Guardian(2021)]%
        {the_guardian_as_2021}
\bibfield{author}{\bibinfo{person}{Guardian}.} \bibinfo{year}{2021}\natexlab{}.
\newblock \bibinfo{title}{As a verdict on the climate crimes of humanity, the
  new {Intergovernmental} {Panel} on {Climate} {Change} report could not be
  clearer: {W}e're guilty as hell.}
\newblock
\newblock
\urldef\tempurl%
\url{https://www.instagram.com/p/CSXHuKIqCs1/}
\showURL{%
\tempurl}
\newblock
\shownote{{S}lide 4. Accessed on: 23 February 2023}.


\bibitem[Hansen and Machin(2008)]%
        {hansen_visually_2008}
\bibfield{author}{\bibinfo{person}{Anders Hansen} {and} \bibinfo{person}{David
  Machin}.} \bibinfo{year}{2008}\natexlab{}.
\newblock \showarticletitle{Visually branding the environment: {Climate} change
  as a marketing opportunity}.
\newblock \bibinfo{journal}{\emph{Discourse Studies}} \bibinfo{volume}{10},
  \bibinfo{number}{6} (\bibinfo{date}{Dec.} \bibinfo{year}{2008}),
  \bibinfo{pages}{777--794}.
\newblock
\showISSN{1461-4456}
\urldef\tempurl%
\url{https://doi.org/10.1177/1461445608098200}
\showDOI{\tempurl}


\bibitem[IPCC(2009)]%
        {ipcc_new_2009}
\bibfield{author}{\bibinfo{person}{IPCC}.} \bibinfo{year}{2009}\natexlab{}.
\newblock \bibinfo{title}{{A} new assessment cycle, a new visual identity}.
\newblock
  \bibinfo{howpublished}{\url{https://wg1.ipcc.ch/sites/default/files/documents/ipcc_visual-identity_guidelines.pdf}
  {Accessed on: 23 February 2023}}.
\newblock


\bibitem[Joslyn and Demnitz(2021)]%
        {joslyn_explaining_2021}
\bibfield{author}{\bibinfo{person}{Susan Joslyn} {and} \bibinfo{person}{Raoni
  Demnitz}.} \bibinfo{year}{2021}\natexlab{}.
\newblock \showarticletitle{Explaining how long {CO2} stays in the atmosphere:
  {Does} it change attitudes toward climate change?}
\newblock \bibinfo{journal}{\emph{Journal of Experimental Psychology: Applied}}
  \bibinfo{volume}{27}, \bibinfo{number}{3} (\bibinfo{date}{Sept.}
  \bibinfo{year}{2021}), \bibinfo{pages}{473--484}.
\newblock
\showISSN{1939-2192, 1076-898X}
\urldef\tempurl%
\url{https://doi.org/10.1037/xap0000347}
\showDOI{\tempurl}


\bibitem[Karer et~al\mbox{.}(2020)]%
        {karer_conceptgraph_2020}
\bibfield{author}{\bibinfo{person}{B. Karer}, \bibinfo{person}{Inga Scheler},
  {and} \bibinfo{person}{H. Leitte}.} \bibinfo{year}{2020}\natexlab{}.
\newblock \showarticletitle{{ConceptGraph}: {A} formal model for interpretation
  and reasoning during visual analysis}.
\newblock \bibinfo{journal}{\emph{Computer Graphics Forum}}
  \bibinfo{volume}{39} (\bibinfo{date}{Feb.} \bibinfo{year}{2020}).
\newblock
\urldef\tempurl%
\url{https://doi.org/10.1111/cgf.13899}
\showDOI{\tempurl}


\bibitem[Kim et~al\mbox{.}(2017)]%
        {kim_bubbleview_2017}
\bibfield{author}{\bibinfo{person}{Nam~Wook Kim}, \bibinfo{person}{Zoya
  Bylinskii}, \bibinfo{person}{Michelle~A. Borkin},
  \bibinfo{person}{Krzysztof~Z. Gajos}, \bibinfo{person}{Aude Oliva},
  \bibinfo{person}{Fredo Durand}, {and} \bibinfo{person}{Hanspeter Pfister}.}
  \bibinfo{year}{2017}\natexlab{}.
\newblock \showarticletitle{{BubbleView}: {An} interface for crowdsourcing
  image importance maps and tracking visual attention}.
\newblock \bibinfo{journal}{\emph{ACM Transactions on Computer-Human
  Interaction}} \bibinfo{volume}{24}, \bibinfo{number}{5} (\bibinfo{date}{Oct.}
  \bibinfo{year}{2017}), \bibinfo{pages}{1--40}.
\newblock
\showISSN{1073-0516, 1557-7325}
\urldef\tempurl%
\url{https://doi.org/10.1145/3131275}
\showDOI{\tempurl}


\bibitem[Lundgard and Satyanarayan(2021)]%
        {lundgard_accessible_2021}
\bibfield{author}{\bibinfo{person}{Alan Lundgard} {and} \bibinfo{person}{Arvind
  Satyanarayan}.} \bibinfo{year}{2021}\natexlab{}.
\newblock \bibinfo{title}{Accessible visualization via natural language
  descriptions: {A} four-level model of semantic content}.
\newblock
\newblock
\urldef\tempurl%
\url{https://doi.org/10.1109/TVCG.2021.3114770/}
\showDOI{\tempurl}


\bibitem[McGrath(2021)]%
        {BBC}
\bibfield{author}{\bibinfo{person}{Matt McGrath}.}
  \bibinfo{year}{2021}\natexlab{}.
\newblock \bibinfo{title}{{Climate} change: {Five things} we have learned from
  the {IPCC} report}.
\newblock
\newblock
\newblock
\shownote{{BBC News at bbc.co.uk/news}
  \url{https://www.bbc.com/news/science-environment-58138714} {Accessed on: 23
  February 2023}}.


\bibitem[Moser(2010)]%
        {moser_communicating_2010}
\bibfield{author}{\bibinfo{person}{Susanne~C. Moser}.}
  \bibinfo{year}{2010}\natexlab{}.
\newblock \showarticletitle{Communicating climate change: {H}istory,
  challenges, process and future directions}.
\newblock \bibinfo{journal}{\emph{WIREs Climate Change}} \bibinfo{volume}{1},
  \bibinfo{number}{1} (\bibinfo{year}{2010}), \bibinfo{pages}{31--53}.
\newblock
\showISSN{1757-7799}
\urldef\tempurl%
\url{https://doi.org/10.1002/wcc.11}
\showDOI{\tempurl}


\bibitem[Nadeem(2021)]%
        {nadeem_gen_2021}
\bibfield{author}{\bibinfo{person}{Reem Nadeem}.}
  \bibinfo{year}{2021}\natexlab{}.
\newblock \bibinfo{title}{Gen {Z}, millennials stand out for climate change
  activism, social media engagement with issue}.
\newblock
\newblock
\urldef\tempurl%
\url{https://www.pewresearch.org/science/2021/05/26/gen-z-millennials-stand-out-for-climate-change-activism-social-media-engagement-with-issue/}
\showURL{%
\tempurl}
\newblock
\shownote{Accessed on: 23 February 2023}.


\bibitem[Naustdalslid(2011)]%
        {naustdalslid_climate_2011}
\bibfield{author}{\bibinfo{person}{Jon Naustdalslid}.}
  \bibinfo{year}{2011}\natexlab{}.
\newblock \showarticletitle{Climate change – the challenge of translating
  scientific knowledge into action}.
\newblock \bibinfo{journal}{\emph{International Journal of Sustainable
  Development \& World Ecology}} \bibinfo{volume}{18}, \bibinfo{number}{3}
  (\bibinfo{date}{June} \bibinfo{year}{2011}), \bibinfo{pages}{243--252}.
\newblock
\showISSN{1350-4509}
\urldef\tempurl%
\url{https://doi.org/10.1080/13504509.2011.572303}
\showDOI{\tempurl}


\bibitem[O'Neill and Smith(2014)]%
        {oneill_climate_2014}
\bibfield{author}{\bibinfo{person}{Saffron~J. O'Neill} {and}
  \bibinfo{person}{Nicholas Smith}.} \bibinfo{year}{2014}\natexlab{}.
\newblock \showarticletitle{Climate change and visual imagery}.
\newblock \bibinfo{journal}{\emph{WIREs Climate Change}} \bibinfo{volume}{5},
  \bibinfo{number}{1} (\bibinfo{date}{Oct.} \bibinfo{year}{2014}),
  \bibinfo{pages}{73--87}.
\newblock
\showISSN{1757-7799}
\urldef\tempurl%
\url{https://doi.org/10.1002/wcc.249}
\showDOI{\tempurl}


\bibitem[Pinker and Feedle(1990)]%
        {pinker_theory_1990}
\bibfield{author}{\bibinfo{person}{Steven Pinker} {and} \bibinfo{person}{R.
  Feedle}.} \bibinfo{year}{1990}\natexlab{}.
\newblock \bibinfo{title}{A theory of graph comprehension}.
\newblock
\newblock
\newblock
\shownote{73--126}.


\bibitem[Robson and McCartan(2017)]%
        {robson2017real}
\bibfield{author}{\bibinfo{person}{Colin Robson} {and} \bibinfo{person}{Kieran
  McCartan}.} \bibinfo{year}{2017}\natexlab{}.
\newblock \bibinfo{title}{Real world research, 4th Edition}.
\newblock
\newblock
\showISBNx{978-1-118-74523-6}


\bibitem[Rothermich et~al\mbox{.}(2021)]%
        {rothermich_influence_2021}
\bibfield{author}{\bibinfo{person}{Kathrin Rothermich},
  \bibinfo{person}{Erika~Katherine Johnson}, \bibinfo{person}{Rachel~Morgan
  Griffith}, {and} \bibinfo{person}{Monica~Marie Beingolea}.}
  \bibinfo{year}{2021}\natexlab{}.
\newblock \showarticletitle{The influence of personality traits on attitudes
  towards climate change – {An} exploratory study}.
\newblock \bibinfo{journal}{\emph{Personality and Individual Differences}}
  \bibinfo{volume}{168} (\bibinfo{date}{Jan.} \bibinfo{year}{2021}),
  \bibinfo{pages}{110304}.
\newblock
\showISSN{0191-8869}
\urldef\tempurl%
\url{https://doi.org/10.1016/j.paid.2020.110304}
\showDOI{\tempurl}


\bibitem[Schneider(2012)]%
        {schneider_climate_2012}
\bibfield{author}{\bibinfo{person}{Birgit Schneider}.}
  \bibinfo{year}{2012}\natexlab{}.
\newblock \showarticletitle{Climate model simulation visualization from a
  visual studies perspective}.
\newblock \bibinfo{journal}{\emph{WIREs Climate Change}} \bibinfo{volume}{3},
  \bibinfo{number}{2} (\bibinfo{date}{March} \bibinfo{year}{2012}),
  \bibinfo{pages}{185--193}.
\newblock
\showISSN{1757-7799}
\urldef\tempurl%
\url{https://doi.org/10.1002/wcc.162}
\showDOI{\tempurl}


\bibitem[Schuster et~al\mbox{.}(2022)]%
        {Schuster2022BeingSO}
\bibfield{author}{\bibinfo{person}{Regina Schuster}, \bibinfo{person}{Laura
  Koesten}, \bibinfo{person}{Kathleen Gregory}, {and} \bibinfo{person}{Torsten
  M{\"o}ller}.} \bibinfo{year}{2022}\natexlab{}.
\newblock \showarticletitle{"{B}eing Simple on Complex Issues" - {A}n Expert
  View on Visual Data Communication of Climate Change}.
\newblock \bibinfo{journal}{\emph{ArXiv}}  \bibinfo{volume}{abs/2211.10254}
  (\bibinfo{year}{2022}).
\newblock


\bibitem[Sheppard et~al\mbox{.}(2008)]%
        {sheppard_can_2008}
\bibfield{author}{\bibinfo{person}{Stephen R.~J. Sheppard},
  \bibinfo{person}{Alison Shaw}, \bibinfo{person}{David Flanders}, {and}
  \bibinfo{person}{Sarah Burch}.} \bibinfo{year}{2008}\natexlab{}.
\newblock \showarticletitle{Can visualisation save the world? - {L}essons for
  landscape architects from visualizing local climate change}.
\newblock \bibinfo{journal}{\emph{Digital Design in Landscape Architecture}}
  (\bibinfo{date}{Jan.} \bibinfo{year}{2008}).
\newblock


\bibitem[Vaismoradi et~al\mbox{.}(2016)]%
        {vaismoradi_theme_2016}
\bibfield{author}{\bibinfo{person}{Mojtaba Vaismoradi},
  \bibinfo{person}{Jacqueline Jones}, \bibinfo{person}{Hannele Turunen}, {and}
  \bibinfo{person}{Sherrill Snelgrove}.} \bibinfo{year}{2016}\natexlab{}.
\newblock \showarticletitle{Theme development in qualitative content analysis
  and thematic analysis}.
\newblock \bibinfo{journal}{\emph{Journal of Nursing Education and Practice}}
  \bibinfo{volume}{6}, \bibinfo{number}{5} (\bibinfo{date}{Jan.}
  \bibinfo{year}{2016}), \bibinfo{pages}{100}.
\newblock
\showISSN{1925-4059}
\urldef\tempurl%
\url{https://doi.org/10.5430/jnep.v6n5p100}
\showDOI{\tempurl}
\newblock
\shownote{Number: 5}.


\bibitem[Xiong et~al\mbox{.}(2021)]%
        {xiong_visual_2021}
\bibfield{author}{\bibinfo{person}{Cindy Xiong}, \bibinfo{person}{Vidya
  Setlur}, \bibinfo{person}{Benjamin Bach}, \bibinfo{person}{Kylie Lin},
  \bibinfo{person}{Eunyee Koh}, {and} \bibinfo{person}{Steven Franconeri}.}
  \bibinfo{year}{2021}\natexlab{}.
\newblock \showarticletitle{Visual arrangements of bar charts influence
  comparisons in viewer takeaways}.
\newblock \bibinfo{journal}{\emph{IEEE Transactions on Visualization and
  Computer Graphics}}  \bibinfo{volume}{PP} (\bibinfo{date}{Sept.}
  \bibinfo{year}{2021}), \bibinfo{pages}{1--1}.
\newblock
\urldef\tempurl%
\url{https://doi.org/10.1109/TVCG.2021.3114823}
\showDOI{\tempurl}


\end{thebibliography}

\end{document}